\documentclass[a4paper,twoside]{article}

\usepackage{epsfig}
\usepackage{subcaption}
\usepackage{calc}
\usepackage{color}
\usepackage{amssymb}
\usepackage{amstext}
\usepackage{amsmath}
\usepackage{amsthm}
\usepackage{multicol}
\usepackage{pslatex}
\usepackage{apalike}
\usepackage{algorithm2e}
\usepackage{comment}
\usepackage{url}
\usepackage[bottom]{footmisc}
\usepackage{SCITEPRESS}     % Please add other packages that you may need BEFORE the SCITEPRESS.sty package.

\begin{document}

\title{Perceptions of Entrepreneurship Among Graduate Students: Challenges, Opportunities, and Cultural Biases}

\author{\authorname{Manuela Andreea Petrescu\sup{1}\orcidAuthor{0000-0002-9537-1466X}, Dan Mircea Suciu\sup{1}\orcidAuthor{0000-0002-5958-419X} }
\affiliation{\sup{1}Babe\c s-Bolyai University, Cluj-Napoca, Romania}
\email{\{manuela.petrescu, dan.suciu\}@ubbcluj.ro}
}


\abstract{The purpose of the paper is to examine the perceptions of entrepreneurship of graduate students enrolled in a digital-oriented entrepreneurship course, focusing on the challenges and opportunities related to starting a business. In today's digital era, businesses heavily depend on tailored software solutions to facilitate their operational processes, foster expansion, and enhance their competitive edge, thus assuming, to a certain degree, the characteristics of software companies. For data gathering, we used online exploratory surveys. The findings indicated that 
%most of the students who participated in this study lacked entrepreneurial experience, and even those who had some, it was about a small part of what it takes to run a business. Although 
although entrepreneurship was considered an attractive option by students, very few of them declared that they intended to start a business soon. The main issues raised by the students were internal traits and external obstacles, such as lack of resources and support. Gender discrimination and cultural biases persist, limiting opportunities and equality for women. In terms of gender, women face limited representation in leadership roles, are expected to do more unpaid 'family work', are perceived as less capable in ding business, and need to prove their skills. Even if women are less discriminated now, both genders agree that women still face discrimination in business domain. In terms of percentages, women mentioned gender discrimination in higher percentages. Addressing these issues requires awareness, education, and policy changes
to ensure fair treatment and opportunities for women.
}

\onecolumn \maketitle \normalsize \setcounter{footnote}{0} \vfill

\section{Introduction}
\label{sec:introduction}
Entrepreneurship is a significant contributor to economic growth, prompting a call for increased education in this field among students, teachers, and employees \cite{KLOFSTEN19}. For this reason, entrepreneurial courses were included in the curricula of many universities and colleges. Various educational workshops and programs were created to expose people to these concepts. European Union has also responded to this growing demand with its \textit{Entrepreneurship 2020 Action Plan}\footnote{\url{https://ec.europa.eu/growth/smes/supporting-entrepreneurship/entrepreneurship-education/ commissions-actions-entrepreneurship-education\_en}}, which aims to foster the exchange of best practices in entrepreneurial education. 

%Therefore, entrepreneurship is encouraged, as it has a great impact on the number of jobs in the market, the overall economic growth, innovation and development. This constitutes the underlying rationale why the dynamic of recently established businesses provides a good metric for the economic perspectives of a society. 
%The European Union oversees the registration of businesses and declarations of bankruptcy through the European Business Statistics (EBS) Regulation in order to track the rates of new company formation and job creation across all countries. %Eurostat \cite{Eurostat} reports that the one-year survival rate for the business economy indicates that approximately 82\% of enterprises established in 2019 remained operational in 2020. Additionally, 

%Eurostat data \cite{Eurostat} suggest a positive correlation between the survival rates of companies and the number of jobs they provide. 
The emphasis on entrepreneurial-centered educational initiatives appears to have yielded positive results, as evidenced by the upward trend in the number of new businesses established in Europe in recent years, up to 2023 \cite{Statista}. However, according to McDowell \cite{McDowell}, entrepreneurs experience a trade-off between dedicating themselves to their own business, which may improve their job satisfaction, and the negative consequences of emotional exhaustion, which tend to outweigh this benefit. Alternatively, \cite{padove2019} found that entrepreneurs report higher levels of family satisfaction. In the contemporary era characterized by digital advances, a wide range of businesses rely heavily on software and digital tactics to flourish and engage in competitive endeavors. %Regardless of whether their core offering is a tangible product, a service, or a combination thereof, software plays a crucial role in facilitating their operational processes, driving their growth strategies, and enabling their capacity to navigate a dynamic and ever-changing business environment. Therefore, 
It can be argued that every entrepreneurial enterprise that embraces innovation can be categorized, to some extent, as a software company \cite{Chandra_2022}.

%In Romania, the software development industry has an increasing impact on GDP (Gross Domestic Product), and year after year, this impact is growing, as the National Institute of Statistics (INSSE) states.
%In the last year, 2022, the impact of the Information Technology (IT) sector in Romania on GDP was 6.9\% \footnote{\url{https://insse.ro/cms/sites/default/files/com\_presa/ com\_pdf/pib\_tr2r2022\_1.pdf}, press release 217/7 September 2022}. When analyzing the data in detail, 64\% of the IT companies have foreign capital, while only 36\% have Romanian capital. Therefore, most of the companies in Romania are subsidiaries of multinational companies. 

%Of the 36\% of companies with Romanian capital, 27\% are small companies (having 2-49 employees) %and 8\% are micro companies (0-1 employees) 
%according to the Employers' Association of the Software and Services Industry in Romania (ANIS). 
Small companies are more innovative; they struggle and can adapt more easily \cite{Guimaraes2017}, so it is essential to stimulate their development %\footnote{Industry Study 2022, Software and ITC Services in Romania, Current Situation and  Outlook in a Local and Global Environment, ANIS, 2022,  \url{https://anis.ro/resurse/}}. 
In this context, our objective is to assess the intention of postgraduate students enrolled in an entrepreneurial course at Babes-Bolyai University to start a business, the challenges they perceive in this particular domain, and cultural biases. This study has the potential to obtain accurate results because it was organized at the university level, so the enrolled students were from 22 faculties within the university, different specializations and different cities in the country.

The present study has been conducted with two specific objectives in mind. The first objective is to find out if there is a correlation between students' perception related to the importance of entrepreneurship and how it affects the decision to start their own business. The second objective is to understand the main factors that discourage recent college graduates from pursuing entrepreneurial endeavors.
%The study is based on information gathered from master program students enrolled in the Fundamentals of Entrepreneurship course during the first semester of the academic year 2022-2023. %The course is elective and is open to students from the 22 faculties of Babes-Bolyai University in Cluj-Napoca. The enrolled students came from the following faculties: Mathematics and Computer Science, Physics, Chemistry, Biology and Geology, Geography, Business, Political, Administrative and Communication Sciences, Psychology and Educational Sciences, Sociology, Law, History and Philosophy, Sociology, Physical Education, Sports, and Theology. 
We took into account and analyzed the responses received from students using a survey related to opportunities and challenges in entrepreneurship. %Their responses were evaluated by applying thematic analysis \cite{Braun19}. % performed by the authors %and automatic text analysis performed by ChatGPT \cite{ChatGPT_0}. 

The paper is organized as follows: Section~\ref{sec:literature_review} covers related work, and in Section~\ref{sec:setup} we provide a thorough explanation of the environment in which we did the research%: the course structure, the characteristics of the participants employed in our study and the methodology we used. 
Specifically, Section~\ref{sec:data_collection} outlines the data collection process and the analysis of our research questions. %, and further, in Section~\ref{sec:ChatGPT}, we compare the results of our analysis with the results exposed by an automatic text analysis tool.
We have also taken measures to address potential threats to the validity of our study, as outlined in Section ~\ref{sec:ThreatsToValidity}. Finally, Section~\ref{sec:Conclusion} contains the conclusions and some thoughts on future work.
%how we see the continuation of our work.

\section{Related Work}
\label{sec:literature_review}

%There exists a significant body of literature pertaining to entrepreneurial pursuits. 
Entrepreneurial pursuits represent a significant part of the specialized literature. One of the primary areas of focus in research refers to the significant obstacles that young people face when embarking on entrepreneurial endeavors. It also assesses the effects of entrepreneurial education in addressing obstacles.

Teaching entrepreneurial skills to students begins with teaching entrepreneurial skills to teachers. The importance of improving teachers' competencies in academic technology (not only in business schools but also in engineering programs) and broadening their pedagogical strategies to incorporate novel and innovative methods for teaching entrepreneurship is emphasized in \cite{DonaldKuratko}. \cite{Pittaway} used a systematic literature review (SLR) to examine several themes in the education of future entrepreneurs. The authors arrive at the conclusion that there exists a lack of clarity regarding the definition of \textit{'entrepreneurship education'} and the specific outcomes that are intended to be achieved through its promotion. Furthermore, it has been recommended that improving the evidence base can be achieved through increased investment in the examination of entrepreneurial education, with the aim of assessing the effectiveness of interventions.

%The article \cite{Graevenitz} provides an analysis of the learning mechanisms utilized in the domain of entrepreneurial education. In this study, the authors examine the impact of entrepreneurial education on a cohort of students. The findings suggest that students' beliefs and perceptions of their entrepreneurial abilities are subject to revision after completing the course, even in cases where they previously lacked interest in entrepreneurship. At the same time, \cite{Couetil} reveals a favorable influence of entrepreneurial education on student intentions, profiles, and entrepreneurial projects, as well as a significant carryover effect on their professional lives. \cite{Liguori} presents a research endeavor in which a sample of 18.000 students from 70 countries and 400 universities responded to a survey on the influence of entrepreneurial educational activities in which they participated. The study highlights potential research inquiries that could be addressed in future research.

Multiple studies, including the aforementioned references and additional sources (\cite{Neck}, \cite{Usmanij}, \cite{Boldureanu}), agree that entrepreneurial education is of considerable significance and has the capacity to influence individuals' mindsets.

The context-dependent and gender-aware impediments for female entrepreneurship are discussed in \cite{EspinozaTrujano2020} and their impact in the digital era in \cite{Ughetto}.

Most papers show that entrepreneurial education is well recognized and needs more research to fully understand its impacts on gender biases. This study empirically examines entrepreneurial perspectives in the context of digital transformation.

\section{Overview and Structure of the Study}
\label{sec:setup}

Our investigation considered the feedback provided by students who were registered in the elective digital-oriented online course "Fundamentals of Entrepreneurship", at Babes-Bolyai University through a concluding survey. This section provides an overview of the course and the demographic of the students.%, and the format of the survey.

\subsection{Course overview}
\label{sec:course_overview}

"Fundamentals of Entrepreneurship" is a cross-disciplinary program accessible to all master's students of Babes-Bolyai University, Romania. The course is elective, contains 14 lectures of two hours long, and is open to students from 22 faculties.

The course was developed with the underlying premise that there is a significant correlation between entrepreneurship, innovation, and software development, which is primarily influenced by market dynamics, scalability, cost effectiveness, and the transformative capabilities of digital technologies. 
We believe that software is an essential enabler of innovation, efficiency, and competitiveness. Some companies are heavily dependent on software, even if they are not classic software companies. Thus, even in non-software businesses, software helps innovate and achieve goals. All of these ideas were considered during the selection process for topics and speaker choices for all course lectures.

The "Fundamentals of Entrepreneurship" course provides students with the opportunity to acquire foundational knowledge in the field of entrepreneurship. This includes instruction on essential skills such as developing business plans, evaluating solutions and understanding the needs of future consumers and competitors. Moreover, numerous topics encompass multidisciplinary elements wherever information technology (IT) assumes a substantial role. Consequently, lecturers frequently discuss topics related to digital transformation and digitization. In addition, the course covered aspects related to learning various software tools, including their optimization and how to create a business plan using them. Each lecture was taught by a specialist in a specific field. The speakers were university colleagues or distinctive individuals from the local entrepreneurial ecosystem. The lectures have a strong emphasis on practical and digitization aspects.

The distinctive character of the course can be attributed to several factors: participant's diversity, teaching method, and speaker's selection. Teaching students from different master programs needs flexible methods to suit varying origins, knowledge levels, interests, and views, enabling cross-disciplinary integration. Due to the geographical distribution of the students, we opted for an online course and online access to course materials, video recordings, evaluations, and announcements. While the course faced organizational challenges due to its diverse participant base, it capitalized on diverse expertise, geographical distribution, and industry insights to create a dynamic and enriching learning environment for technology entrepreneurship students.

At the beginning of the course, we collected information on students' perceptions of gender discrimination and cultural biases. Furthermore, after each lecture, we requested feedback to assess the level of student engagement, their comprehension of the material covered, and their general attitude and perception toward the course and the information presented. 
The authors' research on course organization problems and student evaluation of course content and structure was published in \cite{easeai22}.

\subsection{Participants demographics} 
\label{sec:participants}
As indicated in the preceding section, the students who participated in the course were registered in different faculties located in various locations (urban areas) where the university had established learning centers. As participation in the course was optional, the course was announced on the faculties' pages, and we did not influence the enrolled students' selection process. A total of 401 students enrolled in the master's entrepreneurship course. The subset of students that finally joined the course formed our survey-participant set. A group of students experienced challenges with time availability, which resulted in their withdrawal from the course at the beginning.

\begin{comment}
\subsection{Survey}
\label{sec:survey_structure}
To collect student perceptions about various entrepreneurial aspects, the survey research method (as defined by the ACM Sigsoft empirical standards for software engineering research \cite{ACM}) was used. Although some closed questions were included to categorize the opinions of the participants, most of the questions were open to gain a more comprehensive understanding of the primary concepts conveyed in the students' responses. Given the context mentioned above, it should be noted that the survey was conducted online. To comply with research ethics, we informed students about the optional and anonymous characteristics of the survey, the purpose of collecting data, and how it will be used. Due to the absence of any restrictions imposed on student participation, the number of students who opted to participate in the study was 75. In absolute terms, we considered that we had a sufficient number of responses for our analysis.
\end{comment}

\section{Data Collection and Analysis}
\label{sec:data_collection}

The main objective of our study was to conduct a qualitative investigation into the student's perceptions regarding the challenges and opportunities related to starting a business venture, as well as their inclination toward initiating a new company.  We projected the goal into two
research questions:\newline

\textbf{RQ1}: \textit{What are the main factors that discourage recent college graduates from pursuing entrepreneurial endeavors?} The primary objective is to identify potential modifications to the content of the Fundamentals of Entrepreneurship course that would effectively cater to the critical requirements of young learners. Consequently, we have conducted an analysis of the factors that serve as disincentives for individuals to pursue private entrepreneurship.\newline
%\textbf{RQ2}: \textit{How does the perception of the importance of entrepreneurship affect the decision of recent college graduates to start their own businesses?}
%Subsequently, an examination was conducted on how students situate themselves with respect to entrepreneurship as a whole and the degree of significance they attribute to entrepreneurship within a broader economic framework. This positioning can provide us with valuable insight into the significance of this matter for adolescent learners and their inclination toward pursuing this path.\newline

\textbf{RQ2}: \textit{Which are the cultural biases that influence entrepreneurial activity?} Subsequently, an examination was conducted on how students place themselves with respect to cultural biases and discrimination as a whole and the degree of significance they attribute to entrepreneurship within a broader economic framework. This positioning can provide us with valuable information on the importance of this issue for adolescent learners and their desire to pursue this path.

We collected the responses using an online survey. 
%We chose to use Google Forms because it is easy to use, intuitive, and has a large market share, thus increasing the chances that people have previously worked with it; answering a survey using Google Forms should not pose difficulty in terms of usability/ or ''how to''. 
The survey remained open for two weeks to allow everyone to respond despite a busy schedule. %However, most of the responses were submitted immediately after the course: 83\% of the total number of responses. 
In this study, we used quantitative methods; specific questionnaire surveys according to the specification of empirical community standards \cite{ACM} and thematic analysis \cite{Braun19} to evaluate the responses to open questions. Questionnaire surveys and thematic analysis were used in other studies related to computer science \cite{redmond13,icsoft22,motogna21}.

To comply with the ACM standards \cite{ACM}, we worked in parallel using the following procedure: collecting the data, performing a brief analysis of the responses, reallocating the answers to other questions (if they were better fitted), determining specific keywords, and grouping them into classes. The classification was verified by the other author, who also made some observations. Both authors analyzed the observations together, decided if and what changes were necessary, and performed them.

\begin{comment}

\begin{itemize}
    \item collecting the data
    \item performing a brief analysis of the answers 
    \item reallocating answers to other questions (if they were better fitted)
    \item determining specific keywords
    \item grouping keywords into classes 
    \item the classification was verified by the other author, who also made some observations. 
    \item both authors analyzed the observations together, decided if changes are necessary, and performed them.
\end{itemize}
\end{comment}   

%When working with thematic analysis, we noticed that 
Some answers contained exactly one keyword, other answers contained more keywords, and there were questions for which we did not have any answers. When we analyzed the results, we decided to use the prevalence appearance of the keywords compared to the total number of answers; therefore, the results obtained are expressed in percentages, but the sum of all percentages was not 100\% (because the answers contained more than one keyword). The questions asked in the survey can be visualized in Table \ref{tab:questions}. %Certain questions may appear similar in nature; however, our objective was to determine and evaluate the underlying intentions by considering multiple questions. For example, Q3 is used as a checkpoint for Q2 (even if there is an intention to start a business, if the plan to start it is 10 years from the current date, the intention is weak and is rather declarative). 
Closed-ended questions, such as those about gender, were used to categorize the participants and facilitate the formulation of conclusions.

\begin{table}[h!]
  \caption{Survey Questions}
  \label{tab:questions}
  {\small {
  \begin{tabular}{|p{0.5cm}|p{6.2cm}|}
\hline
  Q1& What faculty are you from? \\
\hline
  Q2& How do you identify yourself? (man, woman, other)\\
\hline
 Q3&  What are in your opinion the major challenges when starting a business?\\ %the possible
\hline
 Q4&  Do you want to start a business? (yes / no) \\
\hline
 Q5&  If you want to start a business, when do you plan to start it? \\%(Linear choice on a scale of 1 to 5, corresponding to a time scale from current future to more than 10 years)\\
\hline
 Q6&  Is entrepreneurship an attractive option for young people? (open question)\\
\hline
 Q7&  Do you believe that women face discrimination in entrepreneurship? (open question) \\
\hline
 Q8&  Do you believe that men face discrimination in entrepreneurship? (open question)\\
%\hline
%Q7.&  How would you describe the contribution of entrepreneurship to the economic development of the country?\\

\hline
\end{tabular}
\vspace{-5mm}
}}
\end{table}

\subsection{RQ1: What are the main factors that discourage recent college graduates from pursuing entrepreneurial endeavors?}

To find the response to this question, we asked students at the beginning of the course what they considered to be the major issues and challenges that could appear when starting a business. From a psychological perspective, people may find it easier to achieve emotional detachment, allowing for a clearer analysis of the challenges faced by young entrepreneurs. A portion of the student population, specifically 14.66\%, opted to abstain from responding to this inquiry. However, other students provided a diverse range of examples that have the potential to influence and dissuade recent college graduates from embarking on entrepreneurial pursuits. 

\subsubsection{General overview}
The identified challenges can be grouped as external challenges (context, laws, other people's influence), and challenges that are tightly related to each person's skills and methods of solving a problem. The most mentioned challenge was an external one, the fight with \textit{general perception} 25.67\%(men are better compared to women or young people can not create a business). This challenge is related to the second most mentioned one: \textit{Lack of respect/authority} mentioned by 9.45\%, as the third: \textit{Lack of credibility} scores half of the second. 
According to the responses received, women also face challenges in terms of credibility due to gender biases. Unfair stereotypes can undermine their expertise and decision-making skills, impacting career opportunities and personal interactions. 

\textit{Fluctuating income} also scores high on perceived challenges. Irregular earnings can lead to financial instability, making it challenging to cover essential expenses consistently. Planning for fluctuating income often involves budgeting, building savings, and establishing a financial safety net to mitigate the potential risks associated with income variability. Other external challenges mentioned by students are \textit{''Fear of failure'' or ''having too high expectations'', ''Time constraints'', ''Family constraints'' (family work, including care of children)}.

The students expressed challenges related to the skills of each person: \textit{Overconfidence, Impulsiveness, Lack of commitment, Perfectionism or Lack of organization}. 
In the students' responses, we could find one or more keywords from both types of challenges: \textit{''Barriers like confidence'', ''Not taken seriously'', ''Gender discrimination, prejudice, sacrifices made in the family'', ''lack of time, perfectionism, responsibility for the family, fear of failure''}. %Some responses reinforced their opinion with statistical information: \textit{''However, women are in a clear minority when it comes to the percentage of women who are CEOs of Fortune 100 companies''}.

%We classified the challenges into three main categories, the first one involves external challenges related to each person (family responsibilities, general biases, and time constraints). The second category includes the skills and characteristics of each person (impulsiveness, fear of failure, etc.). The last category is related to the generic business environment: financial, fluctuating income and competition, as mentioned in Table \ref{tab:challenges}.
We classified the challenges into three main categories, the first one involves external challenges, the second one includes the skills and characteristics of each person. The last category is related to the generic business environment %: financial, fluctuating income and competition,
as mentioned in Table \ref{tab:challenges}.

\begin{table}[h!]
  \caption{Entrepreneurial challenges classification}
  \label{tab:challenges}
  {\small{
  \begin{tabular}{|p{1.5cm}|p{5.0cm}|}
\hline
External challenges & General Perception (Cultural Biases), Gender Discrimination, Lack of respect and authority, Time constraints, Lack of credibility, Family work\\
\hline
Personal challenges & Overconfidence, Impulsiveness, Fear of failure and too high expectations, Lack of skills, Lack of organization, Perfectionism, Lack of commitment \\
\hline
Other  & Fluctuating income, Financial issues, Market competition \\
\hline
\end{tabular}
}}
\vspace{-3mm}
\end{table}

Challenges related to economic skills, marketing, manufacturing, financial knowledge, and lack of resources (human/other types of resources) have an insignificant prevalence in received answers. We found the reason for this when we checked the answers related to the student's previous experience in entrepreneurship - most of them do not have any experience, or they have experience only in a specific area (such as sales). As we collected data at the beginning of the course, our students did not know or perceive these challenges as such. In subsidiaries, it reflects the lack of entrepreneurial knowledge and the importance of teaching entrepreneurship.

In conclusion, the main challenges mentioned by the students refer to external factors, to cultural biases, when the role of women in the family is primarily related to the raising of children.
Subsidiary, they also mention challenges that are personal characteristics: impulsiveness, capacity to make fast decisions, perfectionism, or lack of organization.

\subsection{RQ2: Which are the cultural biases that influence entrepreneurial activity?}
Since companies owned by women are underrepresented, we were interested in the cultural biases related to gender discrimination. As innovation is driven by young people, we analyzed cultural biases related to young people in entrepreneurship.

\subsection{Gender discrimination}
According to INSSE, in Romania, only 26\% of the IT companies were started by women. Taking into account that the number of women in Romania is 5\% larger than the number of men\footnote{\url{https://insse.ro/cms/demography-in-europe/bloc-1b.html?lang=en}}, we can conclude that women are underrepresented. Therefore, it is important to understand the reasons behind this disparity. Knowing the reasons, universities and other institutions could take action to improve the percentage of women in the IT entrepreneurship sector.

When we analyzed the responses, we realized that our participants (regardless of their stated gender) have the same beliefs about young people being perceived as less capable compared to a more mature generation. However, in terms of perception of women versus men's capabilities, the ideas are a little more nuanced: women still struggle with the general perception: \textit{''traditionally, she takes care of the house, family and does not do business''}. Another answer stated the general perception as a challenge for women without offering context or explanations: \textit{'' I would say they ('men') are less negatively discriminated compared to women''}. Both genders seem preoccupied with work/life balance and mention the same issues, one man wrote: \textit{''Balance family and business responsibilities''}, other woman specified: \textit{''Gender discrimination, prejudice, sacrifices made in the family''}.
% A woman's response reflects gender biases: women are seen as the main caregivers of the family, responsible for their well-being: \textit{'' Work-life balance: for many women, it is difficult to maintain a work-life balance, especially when they have family responsibilities''}.% or \textit{''Gender discrimination, prejudice, sacrifices made in the family''}. 
%According to the perceptions of the students, women face more external challenges than men when it comes to becoming successful entrepreneurs. 

The perception of discrimination in entrepreneurship differs by gender as a percentage, although both genders perceive that there is gender-based discrimination. In addition, there is a collective perception in the responses received according to which women are perceived to be less capable than men and have more challenges related to family care, since they are the main caregivers. %Although caregiving can be a rewarding role, it often comes at a cost. %This expectation can place a significant burden on them, affecting their career opportunities and personal development. 
Women often face unique challenges related to time management and involvement due to societal expectations.

Men and women (as genders) are perceived to have different sets of skills that would help them in a specific situation. Men are seen to be more competitive and willing to take risks compared to women: \textit{''Sometimes men risk too much''} and women are seen to have less credibility: \textit{''Women may have less credibility''}. 
There was a small percentage of women, less than 10\%, and less than 5\% of men who considered women not discriminated against. %A woman stated that \textit{''Women are not discriminated, they have won their position''}.%, another mentioned that \textit{''In the XXI century and in a modern Europe, I don't think women are discriminated''}. 
Men seemed to be more straightforward: \textit{''They are not'', ''I don't think they (women) are discriminated''}, even if we got some answers that have more details: \textit{''I don't believe there are disadvantages based on gender or that women are discriminated; If you are capable and determined, gender doesn't matter.''}.

Based on the answers, we could determine a perception of skills that defines genders, \textit{''women are more rigorous''} or \textit{''men are more capable''}, but to determine a pattern, we need a larger number of responses and we need responses to questions related to perception of gender-specific skills. These comments appeared in the answers to questions related to gender discrimination; therefore, we consider that we cannot provide statistics on gender-perceived skills.

\begin{figure}[!htbp]
    
    \includegraphics[width = \linewidth]{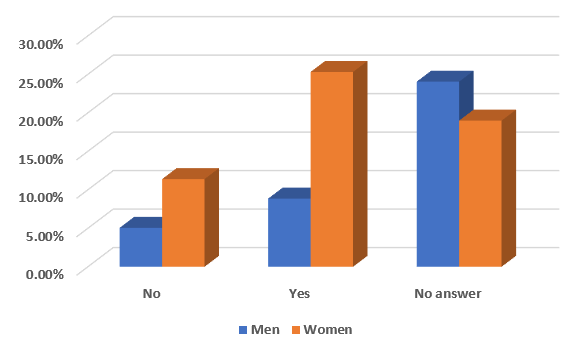}
    \caption{Are women discriminated in entrepreneurship?}
    \label{fig:wdiscrimination}
 
\end{figure}

The percentages of responses considering that women are discriminated are much higher: around 25\% of women and around double for men, as can be seen in Figure \ref{fig:wdiscrimination}. Visualizing, in terms of discrimination, men and women state that women are discriminated in entrepreneurship according to their gender, the percentages are shown in Figure \ref{fig:wdiscrimination}.

\begin{comment}
\begin{figure}[!htbp]
    \centering
    \begin{subfigure}{0.5\textwidth}
        \centering
        \includegraphics[width=\linewidth]{Figures/MenDiscrimination.png} 
        \caption{Perceived men discrimination in entrepreneurship}
        \label{fig:attractive}
    \end{subfigure}
    \hfill
    \begin{subfigure}{0.5\textwidth}
        \centering
        \includegraphics[width=\linewidth]{Figures/WoMenDiscrimination.png} 
        \caption{Perceived women discrimination in entrepreneurship}
        \label{fig:when}
    \end{subfigure}
    \caption{Perceived gender discrimination in entrepreneurship}
    \label{fig:discrimination}
\end{figure}
\end{comment}

\begin{figure}[!htbp]
    \centering
    \includegraphics[width = \linewidth]{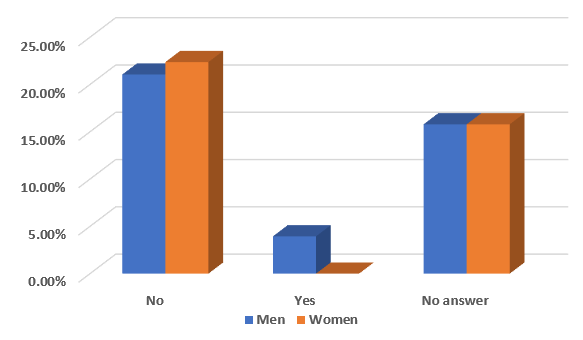}
    \caption{Are men discriminated in entrepreneurship?}
    \label{fig:mdiscrimination}
\end{figure}

However, their open questions reflect a cultural bias, as women are perceived to be less capable compared to men. When asked if women are discriminated, the responses reflected two positions: assumed positions by women: \textit{''lately, I think not'', ''I believe that in our country women are discriminated''} and by men: \textit{''Yes, until they manage to prove their skills. They start their journey with this minor handicap''}. Sometimes, the students' responses reflect the general perception of society: \textit{''Because of the general perception, men are perceived as more capable/resourceful in general''}, \textit{''The mentality is that men are the ones who lead''} or \textit{''women are not taken seriously'', ''misoginism''}. % The idea of cultural discrimination appeared when participants were asked to state if men are discriminated: \textit{''I am not sure, but I would say that men are less negatively discriminated against women.''} - man.
In terms of discrimination, both genders consider that men are less discriminated based on their gender in entrepreneurship, the percentages are shown in Figure \ref{fig:mdiscrimination}.

Some comments stated that general perceptions and cultural biases are changing: \textit{''Unfortunately, sometimes yes, but they (women) are less and less discriminated against, and the world is progressing in favor of equality''}. 
Awareness campaigns and education have empowered individuals to challenge gender biases. %Continuous advocacy, workplace diversity initiatives, and inclusive policies are vital components of progress for women in the modern workforce, and they begin to have effects and raise awareness in society. A man mentioned that {''The promotion of women is sought''}.
While progress has been made, challenges remain, such as the gender gap and underrepresentation in certain fields such as computer science or entrepreneurship.

In conclusion, even if there are opinions that state there are no major challenges related to gender: \textit{''I don't think there are gender disadvantages, if you are capable and determined, it doesn't matter gender''}, the main perception (over 63\% of the answers) mentions that women have more challenges compared to men when we talk about starting and running a business. According to our data, even if both genders state that women are discriminated, women perceive gender discrimination much more compared to men, more than 25\% of women considered to be discriminated versus less than 10\% men.

%Work balance, career, family, and personal life can be challenging. This imbalance does not help women's professional development and limits their involvement in extracurricular activities, community engagement, or personal pursuits. Cultural biases and general perceptions pose other challenges that women must overcome. 

\subsection{Age Influence}

To find the answer to this research question, we analyzed the responses received to the following questions and correlated the answers:
\begin{itemize}
    \item Do you consider entrepreneurship to be an attractive option?
    \item Do you intend to start a business? 
    \item When do you plan to start a business?
\end{itemize}
%Students perceive entrepreneurship as an attractive option for young people. Individuals may be interested in entrepreneurship because they want autonomy and flexibility. Some might like the opportunity to create something new and have a positive impact on their community or industry. 
The large majority of the students (36.00\%) found entrepreneurship very attractive, 34.67\% found it attractive, and 6.67\% of the students did not answer this question. % Taking into account that in the "Fundamentals in Entrepreneurship" course students were mainly young people with a desire to learn business knowledge, the total percentage of students 64.67\% considering entrepreneurship attractive and very attractive can be correlated with their interest in this optional course.

\begin{figure}[!htbp]
    \centering
    \includegraphics[width = \linewidth]{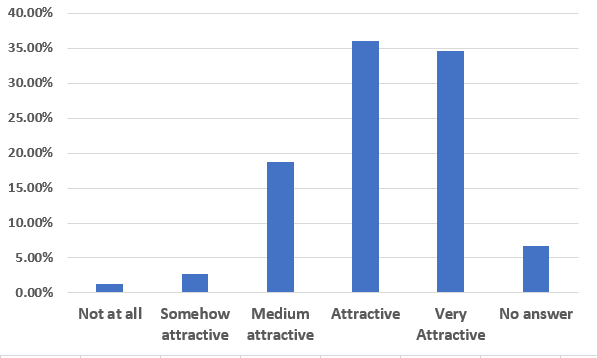}
     \caption{Is entrepreneurship an attractive option?}
        \label{fig:attractive}
\end{figure}

\begin{comment}
\begin{figure}[!htbp]
    \centering
    \begin{subfigure}{0.5\textwidth}
        \centering
        \includegraphics[width=\linewidth]{Figures/IsAttractive.png} 
        \caption{Is entrepreneurship an attractive option for young people?}
        \label{fig:attractive}
    \end{subfigure}
    \hfill
    \begin{subfigure}{0.5\textwidth}
        \centering
        \includegraphics[width=\linewidth]{Figures/WhenToStart.png} 
        \caption{When do the students plan to open a business?}
        \label{fig:when}
    \end{subfigure}
    \caption{Appeal of Entrepreneurship among Young People}
    \label{fig:fullfig}
\end{figure}
\end{comment}
To establish a correlation between the questions, we took into account the distinction between the enticement of entrepreneurship and the inclination to initiate a business venture. %We posed distinct questions to determine the degree to which attractiveness is translated into entrepreneurial intentions. 
Even if entrepreneurship scores high in attractiveness, in our study we found that their intention to start a business is much lower; 45.33\%  stated their interest in opening a business, 44.00\% mentioned that they do not want to start a business, and 10.67\% did not answer this question. Thus, there is a 20\% difference between the stated domain attractiveness and their intentions to start a business. The survey was completed at the beginning of the course, so students lack awareness of potential challenges in various business-related domains, such as financial and competition-related issues, marketing, or developing a product. These challenges will be discovered throughout the course and can have an impact on the percentages. The attractiveness percentages are shown in Figure \ref{fig:attractive}. 

\begin{figure}[!htbp]
    \centering
         \includegraphics[width=\linewidth]{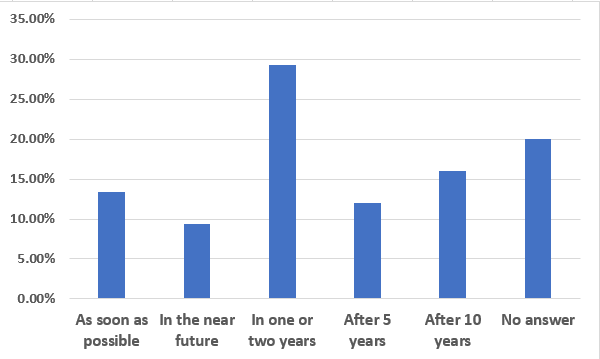} 
        \caption{When do the students plan to start a business?}
        \label{fig:when}
\end{figure}

We used a five-point scale to assess the desire of students to start a business within a designated timeframe. % It is important to note that a stated intention to defer starting a business for a duration exceeding one or two years signifies a relatively weak inclination to initiate such an endeavor. 
We considered that the first two options, \textit{as soon as possible} and \textit{in the near future} are the best predictors of their intentions to start a business. The question was optional, so we considered that the students who did not answer this question do not have specific plans and, most probably, do not have any intentions to start a business soon. When asked in detail to confirm their intention to start a business, a large part of the students stated that they intend to start a business after 5 or more years. Approximately 20\% of the students did not answer this question; we interpreted that the lack of response was due to the fact that these students probably do not want to start a business at all. The percentages are detailed in Figure \ref{fig:when}.

There were answers reflecting a possible reason why students do not want to start a business: cultural bias and perception that consider young people are discriminated: \textit{''yes, if the person is young''}.

In conclusion, students consider entrepreneurship to be an attractive option for young people even if they need to confront some cultural biases, and only 22.66\% stated that they intend to start a business soon or in the near future.

\section{Mitigation Actions to Overcome Challenges}
Depending on the prevalence and type, for each challenge, there should be a different mitigation action. Some challenges such as legal constraints or changes in the financial environment can be mitigated only by learning: Students should realize that there is no perfect /stated solution and that they must adapt depending on the new conditions.

Regarding \textbf{internal challenges}, students face various challenges in their personal skills that can hold them back from reaching their full potential. These challenges come in different forms, such as communication difficulties, time management problems, lack of self-confidence, personal traits (embracing risks more easily), or difficulty in handling stress. However, with self-awareness and acceptance and a willingness to change and adapt, it is possible to overcome these challenges and develop into more capable and resilient individuals. As these challenges are specific to each individual, only a generic approach can be discussed in a course.

The \textbf{external challenges} can be grouped into two major types: those that apply to both genders and those that apply mainly to a specific gender. By cultivating a mindset of perpetual education, students can effectively mitigate various challenges (market understanding, risk management, lack of financial literacy).% they encounter along their entrepreneurial path. 
 Other challenges (such as team building and leadership) %require more than literacy, they
require hands-on actions that can be obtained by working in the domain for other companies or starting a business.

\textbf{Gender biases} have long been an issue in society and impact various aspects of life, including the workplace. These biases perpetuate stereotypes and hinder women's progress. %However, learning and education play a crucial role in challenging and overcoming gender biases. 
By raising awareness, fostering empathy, and promoting inclusivity, individuals can create a more equitable and supportive environment. %However, women need more: they need examples of other women who have succeeded in this domain. %Changing the perceptions of a large percentage of the population is not an easy task, but simply by 
Having more women in this domain, the general perception will change as people adapt to new realities.% So, a really important step is to change the women's perception of themselves and about the things they can achieve.

\section{Threats to Validity}
\label{sec:ThreatsToValidity}
When we developed this study, we considered the guidelines and recommendations for survey research detailed in \cite{ACM}, and we considered that we must address the following threats to validity: the set of participants (target population and participant selection), contingency actions for dropouts, author's subjective approach, and research ethics.

\textbf{Participant set} The survey was sent to all students enrolled in the entrepreneurship course. We did not select any participants; survey-related information was sent to all students enrolled in the course. %Even if we collected data related to the declared gender, we did not modify the participant set to balance participation based on the participant's gender. %Having more men in the course, reflects a greater interest in entrepreneurship. 
Participation in the study was optional and was each student's personal decision. Due to this approach, there was no threat related to the target population and participant selection. 

\textbf{Dropout rates} %are more important when referring to surveys that span over a longer period or consist of multiple steps. 
We tried to mitigate this threat by sending only one online survey, by limiting the number of questions and keeping it open for two weeks.
%so that the students who started completing the survey could complete it in a short time. % We also announced the short amount of time (around 5 minutes) that we estimated a person needs to complete the survey to increase participation and decrease dropout rates. 

Regarding the \textbf{research ethics}, we did not enforce participation (this can be confirmed by the participation rates) and we let everyone know about the anonymous character of the survey. 
We did not put any restrictions on answering all the questions, as the questions were optional. %Thus, not only that students had the option to participate or not in the survey, but they could choose which questions to answer.
Because of this approach, we eliminated biases related to compulsory behavior and ensured that we got relevant data. %The only downside related to this approach was that a higher number of \textit{''Did not respond''} results. Correlated with the number of participants, consider it not to be a validity threat.
We also informed the students about the purpose of collecting these data and how we will use the data collected. 

When humans intervene in a process, there is always the issue of a \textbf{subjective approach} that influences the results. We tried to mitigate this risk by following the guidelines and procedures recommended and used by the computer science community for this type of research and for text interpretation.

\section{Conclusions and Future Work}
\label{sec:Conclusion}

We examined potential challenges to launching a business and opportunities by examining the attitudes and opinions of graduated students about entrepreneurship. Few students express a desire to pursue entrepreneurship in the near future, despite the fact that most find it an attractive alternative. The results emphasize the need for entrepreneurial education to provide a more thorough understanding of the skills and knowledge necessary to successfully establish and operate a business. The study also shows that personal and external traits (such as lack of capital and support) can pose serious obstacles.% for budding entrepreneurs. 

Gender discrimination and cultural biases persist, limiting opportunities and equality for women. 
In terms of gender, women appear to have more challenges compared to men, women face limited representation in leadership roles, and are expected to do more unpaid ''family work'', taking care of children and family. They have to face the stereotypes of society, are seen as less capable, and need to prove their skills, to prove that the general mentality is wrong and that not only men are leaders. %Women can also be great leaders. 
%Even if women are less discriminated now, both genders agree that women still face discrimination. 
In terms of percentages, women mentioned gender discrimination in higher percentages. Addressing these issues requires awareness, education, and policy changes to ensure fair treatment and opportunities for women.

To enhance and motivate students to engage in entrepreneurial endeavors, it is imperative that future research investigates pragmatic solutions to these challenges, as well as techniques aimed at improving entrepreneurial education and mitigating gender prejudice. The fight against gender discrimination is far from over, but there is a positive shift toward a more equitable and inclusive world where individuals are judged based on their abilities rather than their gender. This inclusive world can become a reality with the help of education and policies that promote equal respect and authority for women.
\\

\bibliographystyle{apalike}
{\small
\bibliography{Example}}

%section*{\uppercase{Appendix}}

% If any, the appendix should appear directly after the references without numbering, and not on a new page. To do so please use the following command: \textit{$\backslash$section*\{APPENDIX\}}

\end{document}